%% file: 1Main.tex
\documentclass[11pt,conference]{IEEEtran}

\usepackage{placeins}
\usepackage{graphicx}
\usepackage[export]{adjustbox}
\usepackage{tikz}
\usepackage{tabularx}
\usepackage{multirow}
\usepackage{siunitx}
\usepackage{enumitem}
\usepackage{amssymb}
\usepackage{tabularray}
\UseTblrLibrary{varwidth}
\usepackage{fancyhdr}
\fancypagestyle{aiiotfooter}{%
  \fancyhf{}%
  \fancyfoot[C]{\footnotesize Accepted for publication in the 2026 IEEE 7th World AI IoT Congress (AIIoT 2026).}%
}
\usepackage{dblfloatfix}

\usepackage[acronym]{glossaries}
\input{Glossary.tex}

\usepackage{tikz}
\usepackage{ellipsis}
\usetikzlibrary{calc}
\usetikzlibrary{decorations.pathreplacing,decorations.markings,shapes.geometric,shapes.symbols}
\definecolor{lava}{rgb}{0.81, 0.06, 0.13}
\definecolor{myblue}{HTML}{B0D7FF}

\usepackage[many]{tcolorbox}
\usepackage{makecell}
\usepackage{xcolor} 
\usepackage{amsmath}
\usepackage{esvect}
\usepackage[caption=false,font=footnotesize]{subfig}
\usepackage[utf8]{inputenc}
\usepackage{newtxtext}
\usepackage{newtxmath}
\usepackage{array}
\usepackage{cleveref}
\usetikzlibrary{positioning}
\usetikzlibrary{shapes.arrows}
\tikzset{
    mybox/.style={rectangle,
        draw,
        rounded corners,
        minimum width=2cm,
        inner sep=5pt,
        align=center,
        minimum height=1cm
    },
    myarrow/.style={draw=black,
        fill=white,
        minimum width=1cm,
        single arrow
    },
    longarrow/.style={draw=none,
        shading=axis,
        left color=white,
        right color=gray!20!white,
        minimum width=1cm,
        single arrow,
        anchor=south
    }
    }
\usepackage[export]{adjustbox}
\usepackage{multirow}
\usepackage{listings}
 \graphicspath{{Figures/}}

\usepackage{cite}

\begin{document}

\author{
  \IEEEauthorblockN{
    Tonia Haikal\IEEEauthorrefmark{1},~\IEEEmembership{Student Member,~IEEE}
    Shereen Ismail\IEEEauthorrefmark{2},~\IEEEmembership{Senior Member,~IEEE}
     Eman Hammad\IEEEauthorrefmark{1}~\IEEEmembership{Senior Member,~IEEE} 
  }
\IEEEauthorblockA{\IEEEauthorrefmark{1} iSTAR Lab at Texas A\&M University, College Station, TX 77840, USA.}
\IEEEauthorblockA{\IEEEauthorrefmark{2} Merit Network, Inc., University of Michigan, Ann Arbor, MI 48108, USA.}
}

\title{Bridging High-Level Intent and Network Execution: Detecting Violations and Intent Drift Through Low-Level Traffic Analysis}

\maketitle
\thispagestyle{aiiotfooter}
\begin{abstract}
Intent-Based Networking (IBN) structures a core management pillar for autonomous 6G networks by translating high-level administrative goals into autonomous configurations, yet a critical validation gap persists between declarative intent and data-plane execution. This paper investigates this gap by formalizing low-level flow headers into standardized 7-tuple vectors, establishing an Internal Low-Level Intent (ILI) telemetry interface. Leveraging an empirical dataset of 100.91 million flow records from a distributed honeynet, we evaluate three administrative policy regimes (Strict, Balanced, and Permissive) across two metrics: Policy Violations ($V$) and Intent Drift ($D$). Our results expose a distinct \textit{Compliance Paradox} where widening policy permissiveness systematically suppresses violation counts, yet underlying operational intent drift remains mostly invariant. This demonstrates that conventional, violation-centric tracking are unreliable. Furthermore, an empirical case study show that ILI metrics structural violations can inform closed-loop orchestrators to dynamically recalculate and enforce low-level rules that maintain high-level operational intent.
\end{abstract}


\IEEEpeerreviewmaketitle

\input{2Introduction}

\input{31_Background_and_Related_Work}

\input{5ApproachResults-NEW}

 \vspace*{0.02in} 
\input{6DiscussionConclusion}
\bibliographystyle{IEEEtran}
\bibliography{references}
\end{document}

%% file: Glossary.tex
\newacronym{Mem}{Mem}{Model Size}
\newacronym{PT}{PT}{Processing Time}

\newacronym{AODV}{AODV}{Adhoc On Demand Vector}
\newacronym{RFE}{RFE}{Recursive Feature Elimination}
\newacronym{MI}{MI}{Mutual Information}
\newacronym{SC}{SC}{Smart Contract}
\newacronym{GA-SVM}{GA-SVM}{ Genetic Algorithm-based Support Vector Machine} 
\newacronym {GA-DT}{GA-DT}{Genetic Algorithm-based Decision Tree}
\newacronym {MLP}{MLP}{Multilayer Perceptron}
\newacronym{KNN}{KNN}{K-Nearest Neighbors}
\newacronym{IPFS}{IPFS}{Interplanetary File System}
\newacronym{CompNB}{CompNB}{Complement NB}
\newacronym{NC}{NC}{Nearest Centroid}
\newacronym{VBFT}{VBFT}{Verifiable Byzantine Fault Tolerance}
\newacronym{BFT}{BFT}{Byzantine Fault Tolerance}
\newacronym{OCE}{OCE}{Ontology Consensus Engine}
\newacronym{VRF}{VRF}{Verifiable Random Function}

\newacronym{TSS}{TSS}{Transmitted Signal Strength}
\newacronym{TS}{TS}{Trust Score}
\newacronym {MN}{MN}{monitor node}
\newacronym {MITM}{MITM}{Man-in-the-Middle}
\newacronym {RSS}{RSS}{Received Signal Strength}
\newacronym {PSR}{PSR}{Packet Sending Rate}
\newacronym {PFR}{PFR}{packet forwarding rate} 
\newacronym {FD}{FD}{forwarding Delay }
\newacronym {ECA}{ECA}{Energy Consumption Amount}
\newacronym  {NF}{NF}{Node Affinity}
\newacronym{GBCRP}{GBCRP}{GAN and Block Chain based
secured Routing Protocol} 
\newacronym{GBDT}{GBDT}{gradient boosting decision tree}

\newacronym{SOA}{SOA}{service-oriented architecture}
\newacronym{RNN}{RNN}{recurrent neural networks}
\newacronym{VVLC}{VVLC}{Vehicular Visible Light Communication}
\newacronym{CM}{CM}{cluster member}
\newacronym{V2X}{V2X}{vehicle to everything communications} 
\newacronym{V2V}{V2V}{vehicle to vehicle} 
\newacronym{LTE}{LTE}{Long-Term Evolution}
\newacronym{DSRC}{DSRC}{dedicated short-range communications}
\newacronym{OWC}{OWC}{optical wireless communication}
\newacronym{FSO}{FSO}{free spcae optical communication}
\newacronym{PPM}{$M$-PPM}{multilevel pulse position modulation}
\newacronym{GPS}{GPS}{Global Positioning System}
\newacronym{DC-OFDM}{DCO-OFDM}{direct-current-optical-\gls{OFDM}}
\newacronym{PAM}{$M$-PAM}{pulse amplitude modulation}
\newacronym{EPPM}{EPPM}{Expurgated pulse position modulation}
\newacronym{PAPR}{PAPR}{peak to average power ratio}
\newacronym{DPPM}{DPPM}{Differential pulse position modulation}
\newacronym{OPPM}{OPPM}{Overlapping pulse position modulation}
\newacronym{VPPM}{VPPM}{variable PPM}
\newacronym{ISI}{ISI}{inter-symbol interference}
\newacronym{MPPM}{MPPM}{Multiple pulse position modulation}
\newacronym{FDGAN}{FDGAN}{Fully Distributed Generative
Adversarial Networks}
\newacronym{GOSS}{GOSS}{Gradient-Based One-Side Sampling} 

\newacronym{EFB}{EFB}{Exclusive Feature Bundling}
\newacronym{VPFT}{VPFT}{Verifiable Byzantine Fault Tolerance} 
\newacronym{IFS}{IFS}{Interplanetary File System}
\newacronym{OSI}{OSI}{Open Systems Interconnection}
\newacronym{DoS}{DoS}{Denial of Service}
\newacronym{NFR}{NFR}{non-functional requirement}
\newacronym{MAC}{MAC}{Media Access Control}
\newacronym{IT}{IT}{Information Technology}
\newacronym{IoT}{IoT}{Internet of Things}
\newacronym{WSN}{WSN}{Wireless Sensor Network}
\newacronym{iid}{iid}{independent and identically distributed}
\newacronym{DDoS}{DDoS}{Distributed denial of service}
\newacronym{IDS}{IDS}{Intrusion Detection System}
\newacronym{BC}{BC}{Blockchain}
\newacronym{BS}{BS}{base station}
\newacronym{CH}{CH}{cluster head}
\newacronym{QoS}{QoS}{quality of service}
\newacronym{ML}{ML}{Machine Learning}
\newacronym{LR}{LR}{Logistic Regression}
\newacronym{NB}{NB}{Naive Bayes}
\newacronym{K-NN}{K-NN}{K-Nearest Neighbors}
\newacronym{SVM}{SVM}{Support Vector Machine}
\newacronym{DT}{DT}{Decision Trees}
\newacronym{ANN}{ANN}{Artificial Neural Network}
\newacronym{RF}{RF}{Random Forests}
\newacronym{DL}{DL}{Deep learning}
\newacronym{RL}{RL}{reinforcement learning}
\newacronym{DRL}{DRL}{deep reinforcement learning}
\newacronym{PoW}{PoW}{Proof-of-Work}
\newacronym{AN}{AN}{aggregating node}

\newacronym{P2P}{P2P}{peer-to-peer}

\newacronym{ECDSA}{ECDSA}{elliptic-curve digital signature algorithm} 
\newacronym{tsp}{tsp}{transactions per second}
\newacronym{CPU}{CPU}{central processing unit}

\newacronym{FL}{FL}{Federated Learning}
\newacronym{PPV}{PPV}{positive prediction value}
\newacronym{ERR}{ERR}{Error rate}

\newacronym{GM}{GM}{geometric Mean}
\newacronym{RMSE}{RMSE}{root mean square error}
\newacronym{NRMSE}{NRMSE}{normalized \gls{RMSE}}
\newacronym{ROC}{ROC}{receiver operating characteristics}
\newacronym{RAM}{RAM}{random access memory}
\newacronym{KB}{KB}{kilobytes}
\newacronym{BW}{BW}{bandwidth}
\newacronym{Acc}{Acc}{classification accuracy}
\newacronym{Pd}{$P_d$}{probability of detection}
\newacronym{Pfa}{$P_{fa}$}{probability of false alarm}

\newacronym{Pmd}{$P_{md}$}{probability of misdetection}
\newacronym{HECC}{HECC}{Hyperel-liptic Curve Cryptography}
\newacronym{SDN}{SDN}{software-defined  networking} 
\newacronym{LSTM}{LSTM}{long short-term memory}
\newacronym{PoA}{PoA}{ Proof of Authority}
\newacronym{PoS}{PoS}{Proof of Stake}
\newacronym{DPoS}{DPoS}{Delegated Proof of Stake}
\newacronym{PoC}{PoC}{Proof of Capacity}
\newacronym{PBFT}{PBFT}{practical Byzantine fault tolerance}

\newacronym{CA}{CA}{Certificate Authority}
\newacronym{AES}{AES}{Advanced Encryption Standard} \newacronym{GA}{GA}{Genetic Algorithm}

\newacronym{CNN}{CNN}{Convolutional Neural Network}
\newacronym{HMM}{HMM}{Hidden Markov Model}
\newacronym{DNN}{DNN}{Deep Neural Network}
\newacronym{GAN}{GAN}{Generative Adversarial Networks}
\newacronym{ADC}{ADC}{Analog to Digital Converter}
\newacronym{PoET}{PoET}{Proof of Elapsed Time}
\newacronym{tps}{tps}{transactions per second} 
\newacronym{CPS}{CPS}{Cyber-Physical Systems}
\newacronym{HGB}{HGB}{Histogram Gradient Boost}
\newacronym{LEACH}{LEACH}{Low Energy Adaptive Clustering Hierarchy protocol}

%% file: 2Introduction.tex
\section{Introduction}
\label{sectionI}

Intent-Based Networking (IBN) fundamentally redefines network management by shifting operational frameworks from manual, device-specific configuration scripts to goal-oriented, declarative autonomy \cite{zeydan2020recent}. Under an IBN framework, network operators/engineers articulate what the network must achieve through high-level declarative statements, that the control plane is responsible for dynamically computing, compiling, and enforcing as low-level infrastructure configurations \cite{leivadeas2022survey}. Significant progress has been achieved in architectures that support intent translation pipelines leveraging policy compilers, orchestration graphs, and large language models (LLM) ingest engines~\cite{mekrache2024intent}. However, critical issues remain unresolved within the intent validation and closed-loop assurance lifecycles~\cite{zacarias2026enhancing}.  

IBN implements a continuous, closed-loop management lifecycle to automate configuration states through a feedback loop of intent ingestion, automated compilation, policy enforcement, and real-time validation. Current assurance frameworks operate primarily at the control plane, verifying that target configurations have been pushed to software-defined tables \cite{khan2021intent}. They lack empirical, data-plane native validation models capable of verifying whether live traffic truly conforms to abstract operational policies, where execution is enforced exclusively through low-level flow headers.

This paper establishes an empirical-based methodology that treats data-plane flow tuples, comprised of source and destination IP addresses, MAC addresses, transport-layer ports, and protocol headers, as the explicit, atomic execution surface of human intent. We formalize these entities as \textit{Internal Low-Level Intents} (ILI). By doing so, we introduce a traceable construct that maps high-level abstract policy directly to empirical packet streams, enabling real-time, flow-level assurance.

Our methodology utilizes large-scale, high-entropy empirical telemetry collected from an operational distributed HoneyTrap-based honeynet at Merit Network to evaluate the operational behavior of the proposed framework~\cite{haikal2025honeynets}. This environment provides a highly dense and unpredictable traffic stream, serving as an ideal testing ground to analyze the operational bounds of intent-to-flow translation under complex, real-world conditions. By mapping three distinct administrative configurations (Strict, Balanced, and Permissive policies) against a comprehensive dataset of 100,913,000 complete-flow records, we systematically evaluate two metrics: \textit{Policy Violations} ($V$) and \textit{Intent Drift} ($D$). 

This data-driven strategy utilizes low-level network identifiers as a flow-level telemetry interface for continuous data-plane assurance, establishing an empirical baseline that remains invariant across shifting administrative configurations. We identify a distinct decoupling where loosening policy constraints suppresses explicit violation counts, yet the underlying operational deviation, quantified via intent drift, remains static. While conventional monitoring relies on reactive, bottom-up firewall adjustments, mapping low-level structural traffic concentrations within an IBN framework provides the empirical foundation to automatically derive semantic, service-aware policy abstractions. These insights enable the transition to active, closed-loop orchestrators capable of dynamically recalculating and enforcing low-level rules that maintain high-level operational intent despite administrative masking, critical for supporting highly dynamic and multi-tenant 6G environments,

The contributions of this work are: (i) formalizing raw headers into 7-tuple vectors to establish an Internal Low-Level Intent (ILI) telemetry interface; (ii) processing 100.91 million real-world flow records to prove that intent drift is policy-invariant; and (iii) demonstrating how structural violation concentrations can provide insights to enable closed-loop orchestrators to dynamically recalculate and enforce low-level rules that maintain high-level operational intent.

%% file: 31_Background_and_Related_Work.tex
\section{Related Work and Background}
\label{section:background}

The closed-loop lifecycle of Intent-Based Networking (IBN) structures an autonomous orchestration process relying on continuous feedback loops spanning intent ingestion, compilation translation, policy enforcement, and telemetry assurance \cite{zeydan2020recent, leivadeas2022survey}. Substantial research has targeted upstream automation within this architecture to optimize how declarative administrative goals map onto active configurations. Automated policy generation platforms compile high-level demands down to software-defined configuration templates \cite{khan2021intent}, while Large Language Model (LLM) pipelines facilitate natural language intent ingestion within network orchestrators \cite{mekrache2024intent}. More recently, neurosymbolic AI frameworks have been investigated to enhance contextual execution and service management intelligence \cite{colombi2025investigating}.

Existing validation systems operate primarily within the control plane via compile-time logic checks, pre-deployment syntax validation, or structural model parsing. While verifying configuration syntax, they lack empirical, data-plane models to assess policy fidelity under active runtime conditions. This decoupling between static validation and live execution introduces systemic liabilities, as automated pipelines can mask latent attack surfaces, compilation loops, and policy ambiguities that manifest exclusively at runtime \cite{kim2024security}.

To evaluate these autonomous network behaviors under realistic conditions, infrastructure-wide intent testing datasets have been generated to map 5G-and-beyond/6G communication traces  \cite{andradehoz2024infrastructure}, standard frameworks treat low-level flow identifiers as static configuration artifacts rather than dynamic, actionable objects of intent. Bridging this gap requires continuous, telemetry-driven assessment engines. Active telemetry loops ensure multi-tenant isolation \cite{velasco2021end}. This paper unifies these perspectives by treating the live 7-tuple data-plane header as the active, executable unit of translated policy, establishing a granular Internal Low-Level Intent (ILI) telemetry interface for continuous data-plane assurance.

Evaluating the performance of an ILI-driven framework is best assessed against high-entropy traffic as benign production or controlled environment traces possess predictable patterns that can mask policy drift. Empirical unsolicited traffic from network telescopes and distributed honeynets provides an ideal high-entropy environment. Such wide-area monitoring captures scanning behaviors, automated multi-vector botnets, darknet trends \cite{years2025merit}. In particular, our evaluation leverages empirical subsets of the massive packet traces captured across distributed HoneyTrap infrastructures, which map real-world adversarial probing behaviors \cite{haikal2025honeynets}. By repurposing this threat telemetry as an adversarial sandbox across alternative policy tiers, we study the limitations of conventional violation metrics.


%% file: 5ApproachResults-NEW.tex
\begin{figure*}[!htp]
    \centering
    \includegraphics[width=\textwidth]{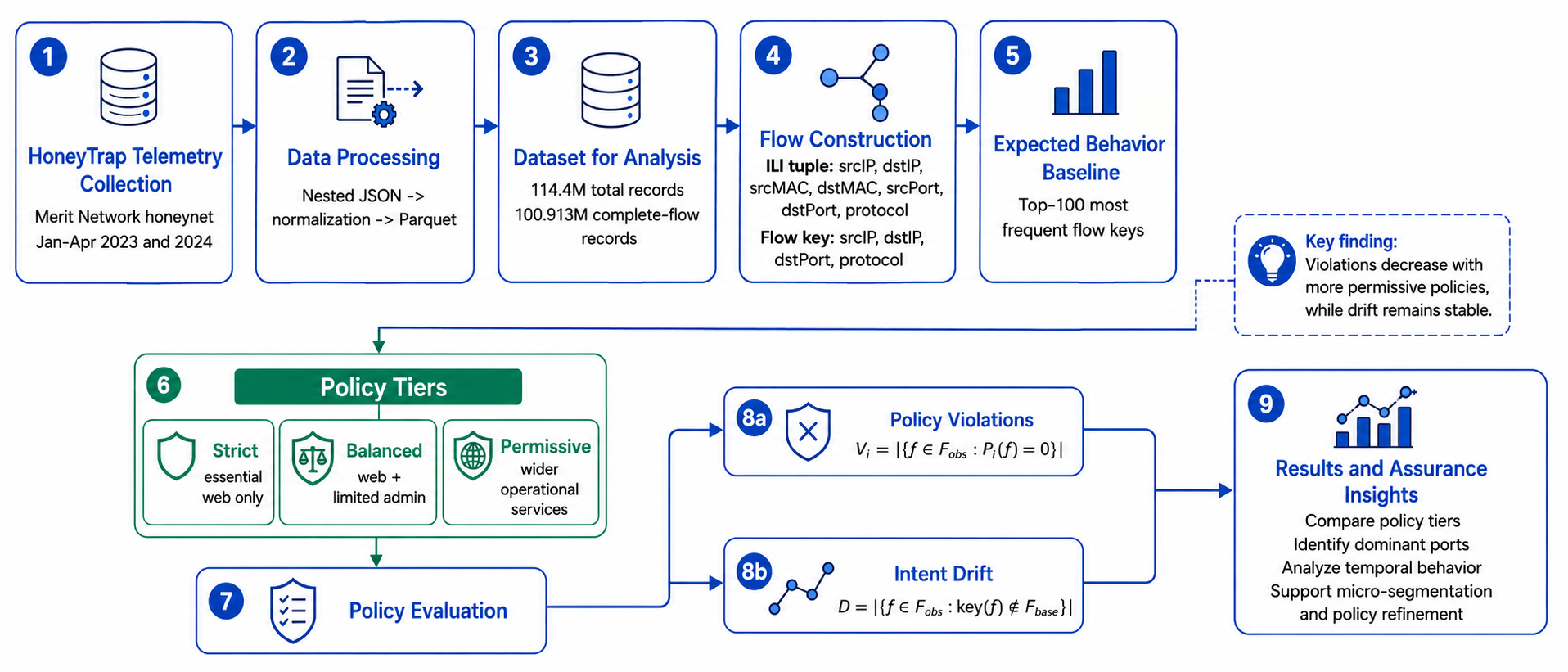}
    \caption{Workflow of the proposed intent-assurance framework.}
    \label{fig:approach_flowchart}
\end{figure*}

\section{Approach and Empirical Results}

Figure~\ref{fig:approach_flowchart} illustrates the architecture of the proposed intent-assurance framework, spanning distributed honeynet telemetry collection, flow construction, policy evaluation, drift measurement, and closed-loop assurance generation. Raw, unsolicited traffic is normalized into structured 7-tuple Internal Low-Level Intent (ILI) vectors to establish an empirical, data-plane native validation layer. By evaluating these observed data-plane primitives against parameterized administrative policy regimes, the framework decouples superficial administrative compliance from underlying operational stability. This provides the critical feedback required for closed-loop intent orchestrators to dynamically recalculate low-level rules and enforce dynamic reconfigurations within autonomous 6G environments.

\subsection{Semantic Intent Data-Plane Formalization}
While traditional IBN closed-loop assurance frameworks verify configuration states at the control plane, they lack mechanisms to validate whether live data-plane traffic conforms to high-level policies. In practice, abstract administrative security and operational objectives are stripped of abstraction at the data plane, where execution is enforced exclusively through discrete, low-level flow headers. We formalizing raw data-plane flow headers into standardized 7-tuple vectors as a granular Internal Low-Level Intent (ILI) telemetry interface for continuous data-plane assurance:
\vspace{-5pt}
\[
\begin{aligned}
f = (&srcIP, dstIP, srcMAC, dstMAC,\\
     &srcPort, dstPort, protocol)
     \vspace{-10pt}
\end{aligned}
\]
\vspace{-5pt}
These vectors serve as the active, executable unit of translated policy.

\subsection{Dataset Flow-Level Telemetry and Vector Projection}
To validate this framework under high-entropy network traffic we leverage a large-scale telemetry corpus collected from a distributed honeynet infrastructure deployed at Merit Network. The raw dataset, structured as nested, time-stratified JSON records, is parsed and normalized into high-performance Parquet repositories. Table~\ref{tab:dataset_summary} summarizes the dataset attributes, isolating a massive evaluation subset of $100,913,000$ flow records.

\begin{table}[htbp]
\centering
\caption{Dataset summary for intent-assurance analysis.}
\label{tab:dataset_summary}
\begin{tabular}{p{0.47\linewidth} p{0.43\linewidth}}
\hline
\textbf{Dataset Attribute} & \textbf{Value} \\
\hline
Raw data format & Nested JSON files  \\
Organization & Year/Month/Day/Hour \\
Processed format & Parquet  \\
Total records & 114.4 million  \\
Complete-flow records & 100,913,000  \\
Flow construction fields & srcIP, dstIP, dstPort, protocol, timestamp \\
Flow key fields & srcIP, dstIP, dstPort, protocol  \\
\hline
\end{tabular}
\vspace{-10pt}
\end{table}

\begin{figure}[htbp]
    \centering
    \includegraphics[width=\linewidth]{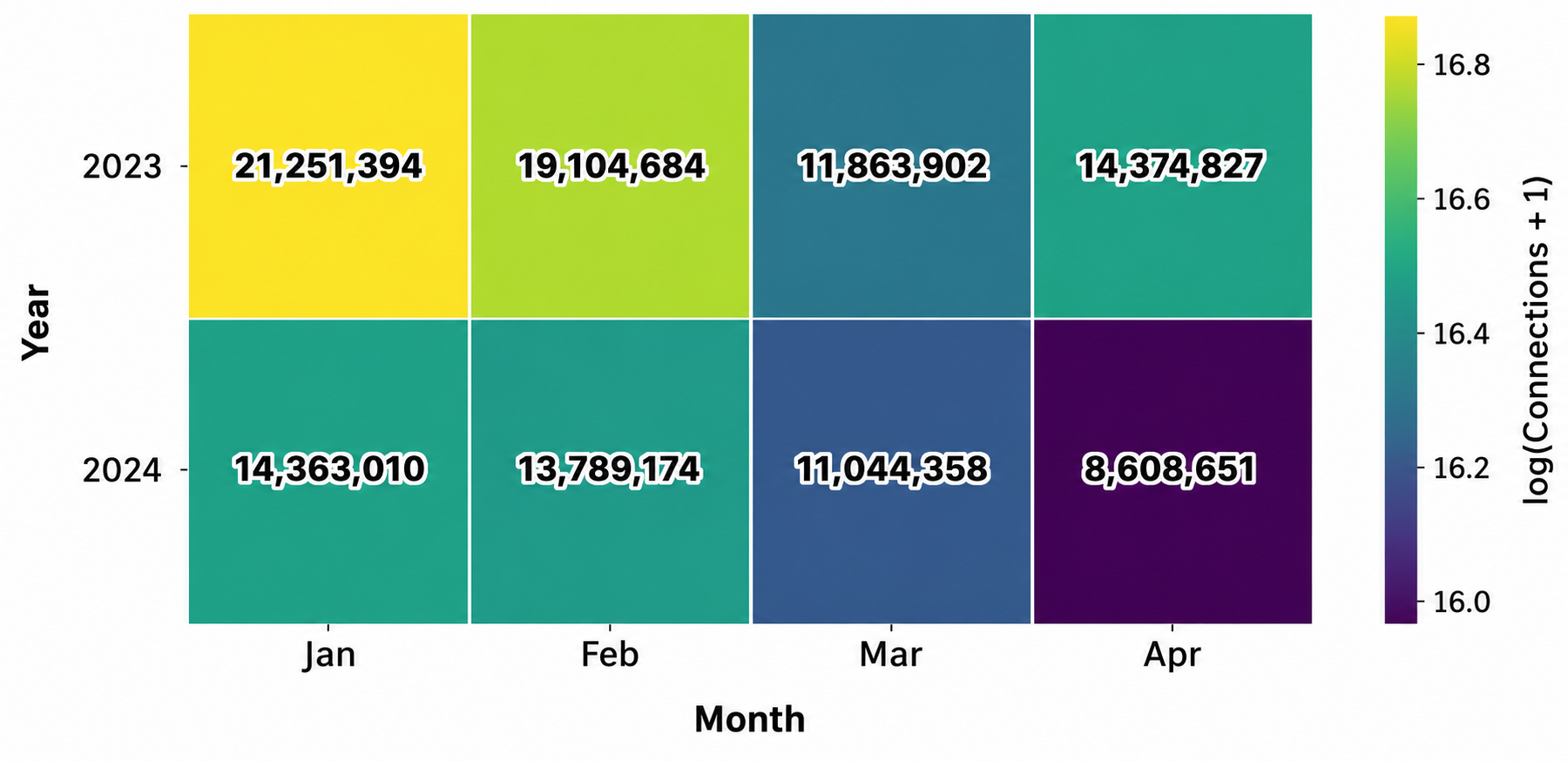}
    \caption{Activity heatmap for January through April for 2023, and 2024.}
    \label{fig:heatmap_intent_policies}
    \vspace{-15pt}
\end{figure}

For each complete-flow record, the framework applies a projection mapping to derive a standardized flow key:
\[
key(f) = (srcIP, dstIP, dstPort, protocol)
\]
This key serves to 1) evaluate data-plane compliance against explicit policy constraints and 2) track behavioral deviations relative to an empirical expected baseline. To establish temporal context, Figure~\ref{fig:heatmap_intent_policies} maps traffic intensity across matching seasonal windows in 2023 and 2024, demonstrating highly concentrated activity in the earlier months of each operational cycle.

\subsection{Data-Driven Policy Construction}
\begin{figure}[htbp]
    \centering
    \includegraphics[width=\linewidth]{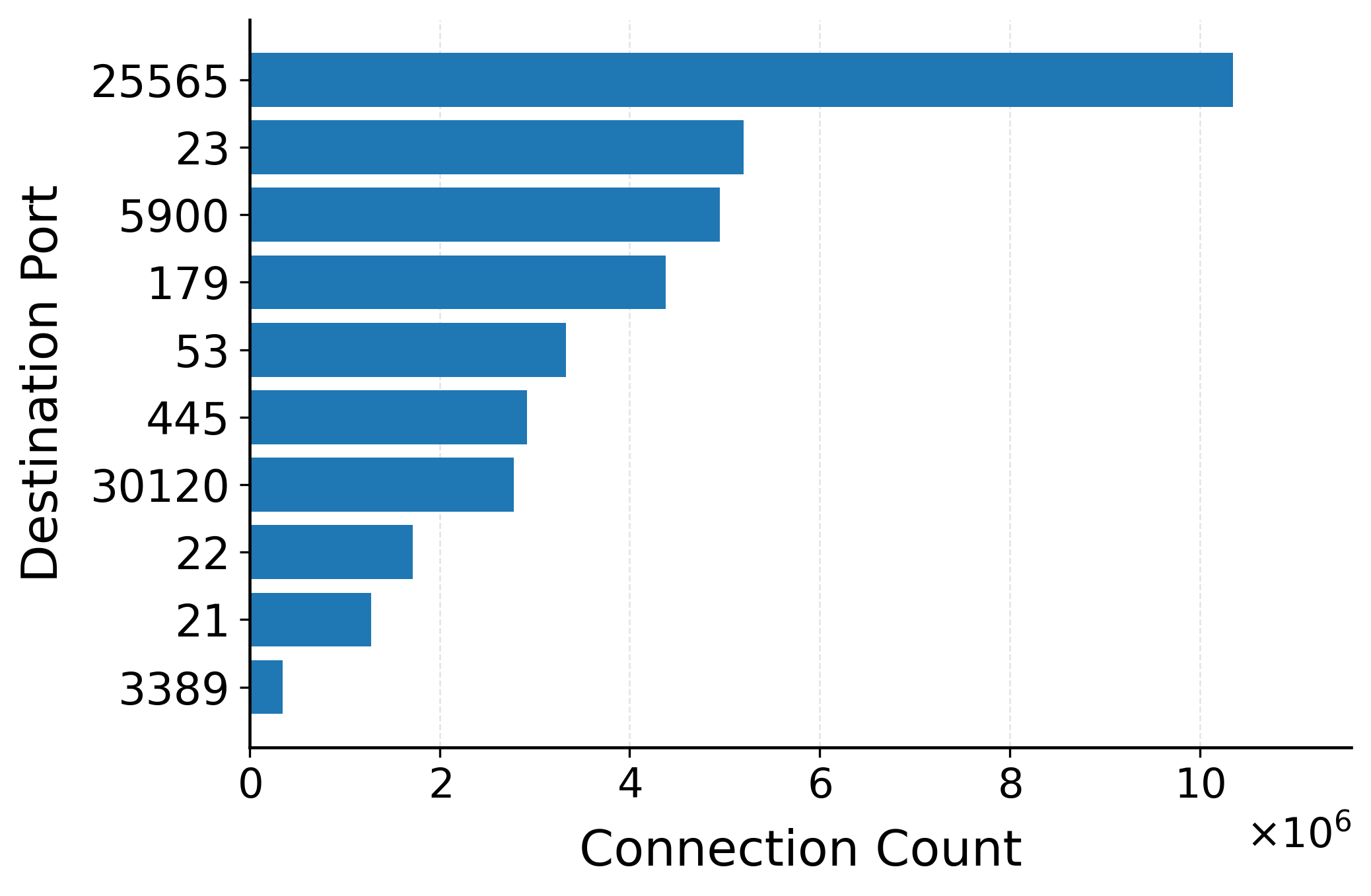}
    \vspace{-10pt}
    \caption{Top Destination Ports by Connection Count.}   
    \label{fig:standout_destination_ports_comparison}
\end{figure}

An IBN Orchestrator must translate abstract, declarative human goals formulated via the Intent Behavioral Language (IBL) into deterministic data-plane enforcement profiles. Rather than relying on arbitrary configurations, we utilize a data-driven approach to identify key service clusters within the network's operational footprint. As illustrated in Figure~\ref{fig:standout_destination_ports_comparison}, initial telemetry analysis reveals that data-plane traffic is not uniformly distributed across the service space; instead, connection volumes cluster heavily around a narrow, predictable subset of destination ports. Port 25565 emerges as a dominant structural outlier, accompanied by distinct administrative and operational vectors including ports 5900 (VNC), 179 (BGP), 22 (SSH), 23 (Telnet), and 445 (SMB). 

Rather than enforcing a coarse, monolithic firewall posture, these empirical concentrations justify service-oriented policy groupings that mirror real-world operational tiers. The IBN Policy Compiler instantiates these behaviors by constructing three nested, mathematically defined administrative policy profiles to evaluate intent enforcement elasticity:
\begin{align*}
    \mathcal{P}_{\text{Strict}} &\implies \{ \text{Essential web services: 80, 443} \} \\
    \mathcal{P}_{\text{Balanced}} &\implies \mathcal{P}_{\text{Strict}} \cup \{ \text{Standard remote admin: 22, 179} \} \\
    \mathcal{P}_{\text{Permissive}} &\implies \mathcal{P}_{\text{Balanced}} \cup {} \\[-2pt]
    &\quad \{ \text{Extended operational services:} \\[-2pt]
    &\qquad \text{21, 23} \}
\end{align*}

where highly sensitive or high-risk ports identified by the telemetry profiling (e.g., 25565, 445) are designated as explicitly restricted across all operational tiers ($\text{port} \in \mathcal{P}_{\text{restricted}}$). This multi-tiered setup allows us to precisely observe how the closed-loop system responds when the formal boundaries of compliance are expanded or contracted.

\subsection{Violations vs. Intent Drift Metrics}
We formalize two data-plane metrics to support assurance: Policy Violation ($V$) and Intent Drift ($D$). Let $F_{obs}$ denote the set of observed flow records. Let $P_i(f) \in \{0, 1\}$ define the policy predicate for a given administrative tier $i$, returning $1$ if $f$ is explicitly authorized and $0$ if restricted. The predicate is evaluated via destination-port membership in the defined service sets. The total violation count $V_i$ under policy tier $i$ is:
\[
V_i = \left| \{ f \in F_{obs} : P_i(f) = 0 \} \right|
\]
This metric captures explicit non-compliance and is highly dependent on the strictness of the active administrative rules. The second metric, Intent Drift ($D$) measures behavioral stability independent of active policy bounds. Let $K$ denote the global set of observed flow keys. We establish an empirical expected-flow baseline $F_{base}$ by isolating the top $k$ dominant recurring communication patterns ($k=100$):
\[
F_{base} = \text{Top}_{k=100}(K)
\]
Then, the total intent drift can be defined as the cardinality of flows whose keys fall outside this baseline:
\[
D = \left| \{ f \in F_{obs} : key(f) \notin F_{base} \} \right|
\]
While $V_i$ fluctuates based on administrative configurations, $F_{base}$ remains fixed, allowing the framework to expose hidden operational deviations that satisfy permissive policies but represent structural behavior drift.

\subsection{The Compliance Paradox}
\begin{figure}[htbp]
    \centering
    \includegraphics[width=\linewidth]{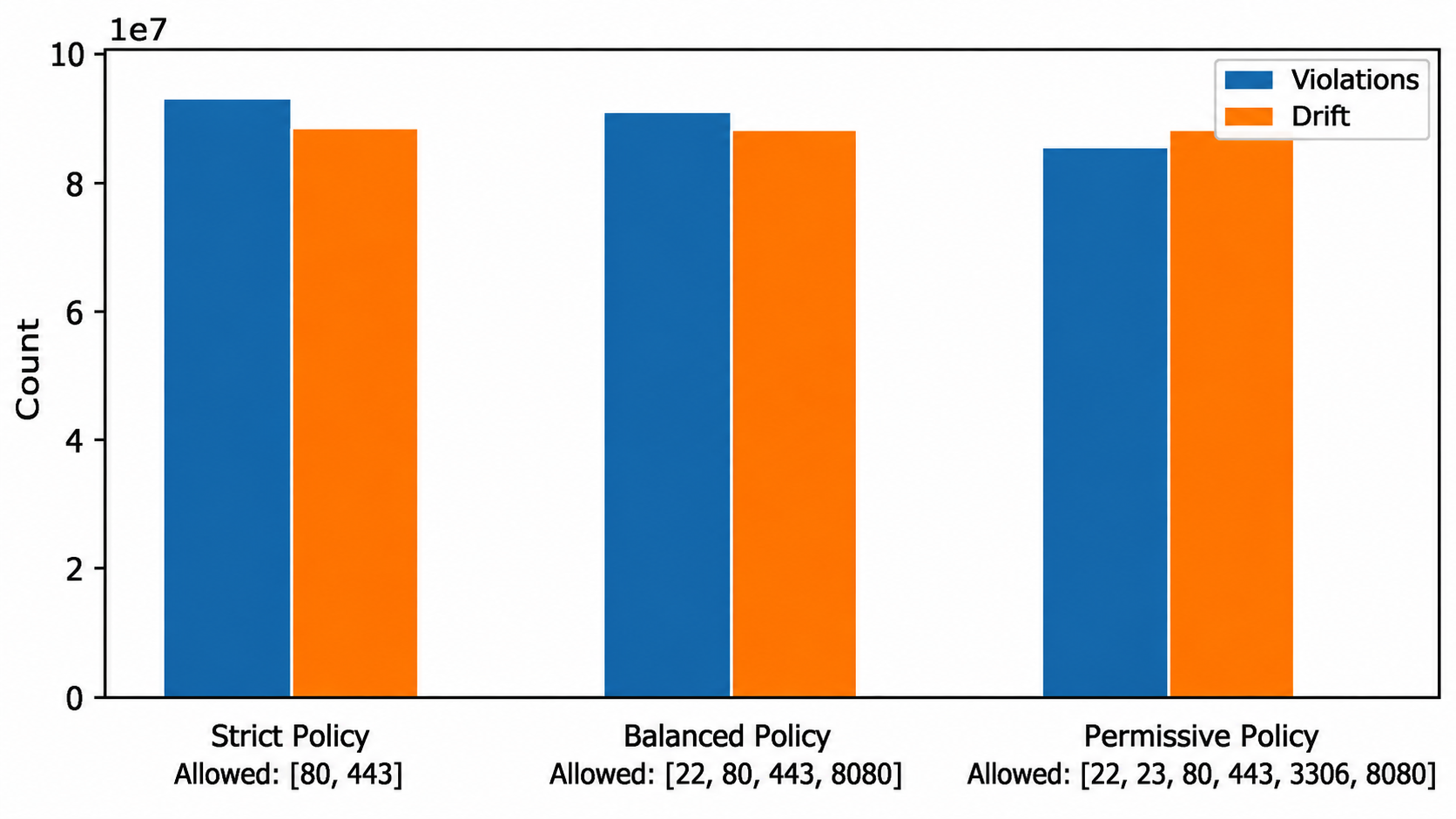}
    \vspace{-10pt}
    \caption{Intent Policy Comparison: Total Violations and Intent Drift.}
    \label{fig:policy_scorecard}
\end{figure}

Evaluating the 100.91 million flow records exposes a fundamental \textit{Compliance Paradox}. As shown in Figure~\ref{fig:policy_scorecard}, total policy violations exhibit a strict monotonic decrease as administrative parameters expand, dropping from 95,024,343 under the Strict tier to 92,988,515 (Balanced) and 87,701,038 (Permissive). However, underlying intent drift remains strictly invariant at exactly 89,031,223 flows across all policy tiers. These results showcase how conventional compliance tracking via violations can be susceptible to administrative masking; widening policy permissiveness suppresses violation alerts without reducing actual data-plane drift.

\begin{figure}[htbp]
    \centering
    \includegraphics[width=\linewidth]{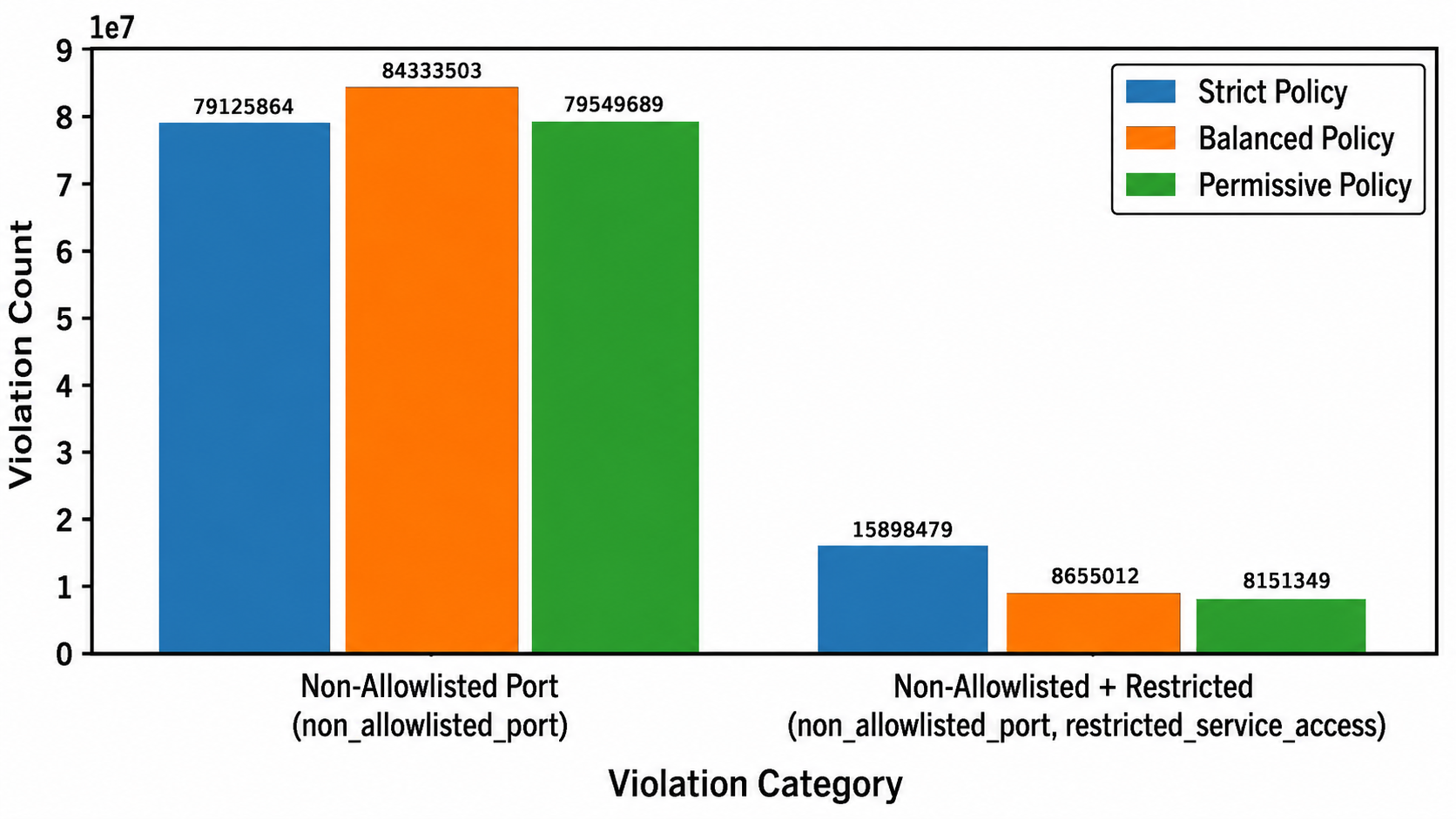}
    \caption{Violation Type Distribution Across Intent Policies.}
    \label{fig:violation_type_distribution}
\end{figure}

This observation is further investigated in Figure~\ref{fig:violation_type_distribution}, which maps the composition of traffic non-compliance across two mutually exclusive tracking categories: 
\begin{itemize}
    \item \textit{Non-Allowlisted Port:} Flows destined to a port that is omitted from the active allowlist ($\text{port} \notin \mathcal{P}_{\text{allowed}}$) but is not flagged as a critical security risk ($\text{port} \notin \mathcal{P}_{\text{restricted}}$).
    \item \textit{Non-Allowlisted + Restricted:} Flows that simultaneously violate allowlist constraints and target highly sensitive, explicitly restricted service ports ($\text{port} \notin \mathcal{P}_{\text{allowed}} \cap \text{port} \in \mathcal{P}_{\text{restricted}}$).
\end{itemize}
Intent assurance would require analyzing the aggregated sum of both classifications. Across the configurations, the true total volume conforms precisely to the policy-strictness hierarchy, moving from 95,024,343 violations under Strict (79,125,864 Non-Allowlisted + 15,898,479 Joint Restricted) down to 92,988,515 under Balanced and 87,701,038 under Permissive. 

The Figure illustrates that shifting policy parameters changes not only the net alert volume but also the underlying mathematical composition of those violations. For example, under the Strict regime where $\mathcal{P}_{\text{allowed}} = \{80, 443\}$, standard administrative ports like Port 22 (SSH) are omitted from the allowlist and belong to the restricted set, classifying their traffic under \textit{Non-Allowlisted + Restricted}. When the orchestrator transitions to the Balanced regime, Port 22 is integrated into the expanded allowlist. Consequently, its traffic instantly ceases to trigger violations, reducing the joint-restricted volume by over 7.2 million flows and altering the visible profile of the chart. Shifting administrative definitions merely reclassifies or masks existing flow behaviors without altering the physical traffic matrix.

\begin{figure}[htbp]
    \centering
    \includegraphics[width=\linewidth]{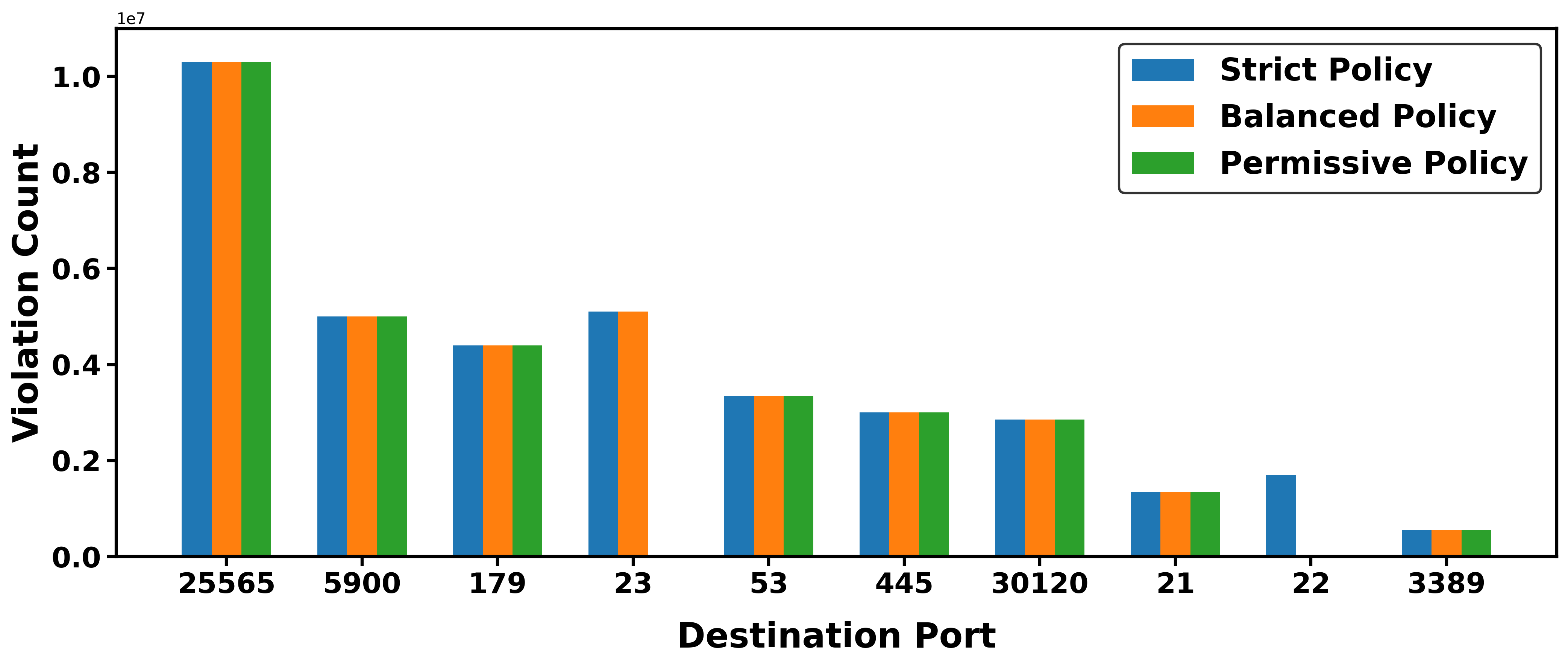}
    \vspace{-10pt}
    \caption{Top Violating Destination Ports Across Intent Policies.}
    \label{fig:top_violating_ports}
    \vspace{-10pt}
\end{figure}

\subsection{Temporal Stability and Policy Consistency}
\begin{figure}[htbp]
    \centering
    \includegraphics[width=\linewidth]{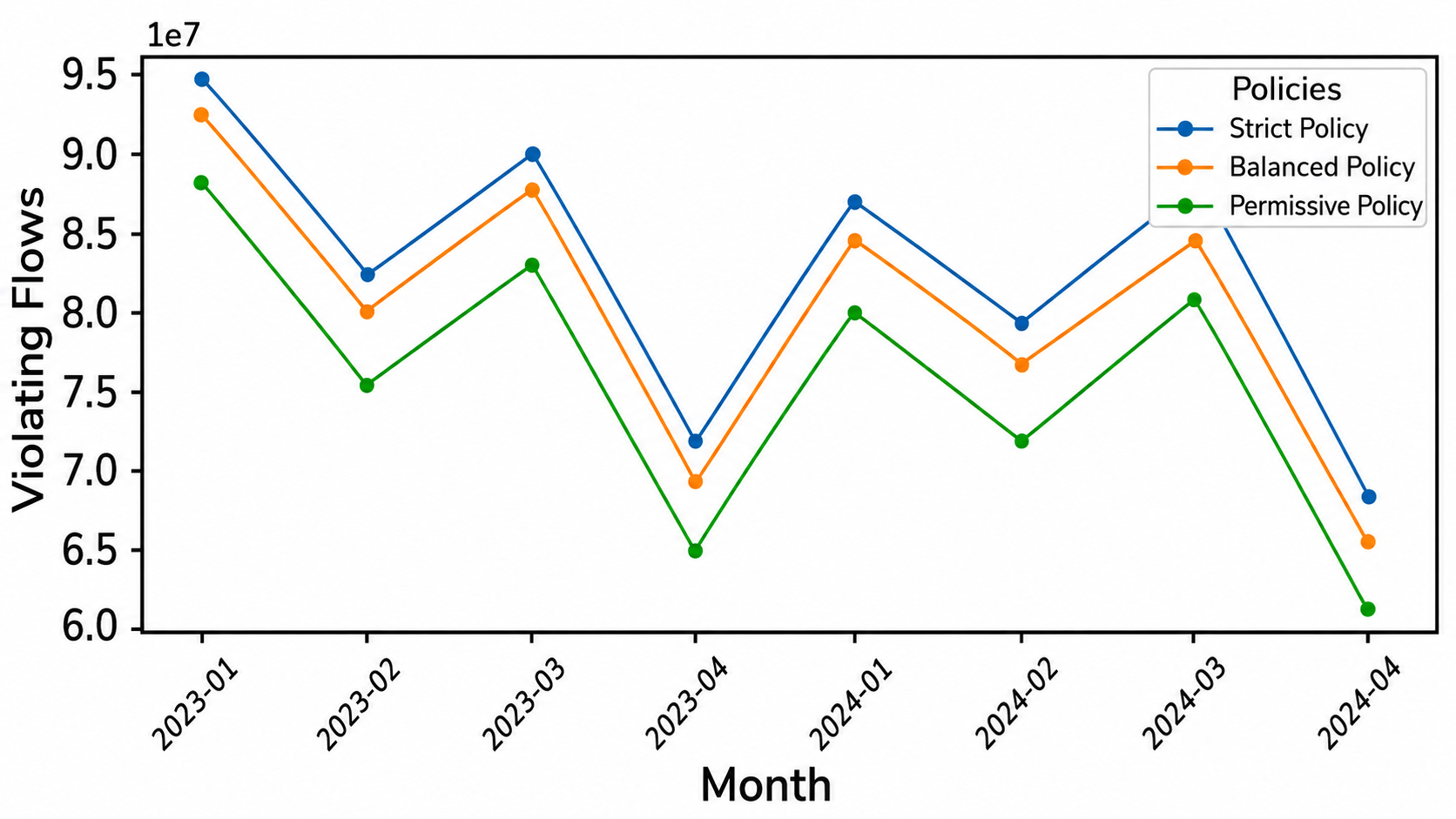}
    \caption{Monthly Policy Violations Across Intent Policies.}
    \label{fig:monthly_policy_violations}
    \vspace{-10pt}
\end{figure}

\begin{figure}[htbp]
    \centering
    \includegraphics[width=0.9\linewidth]{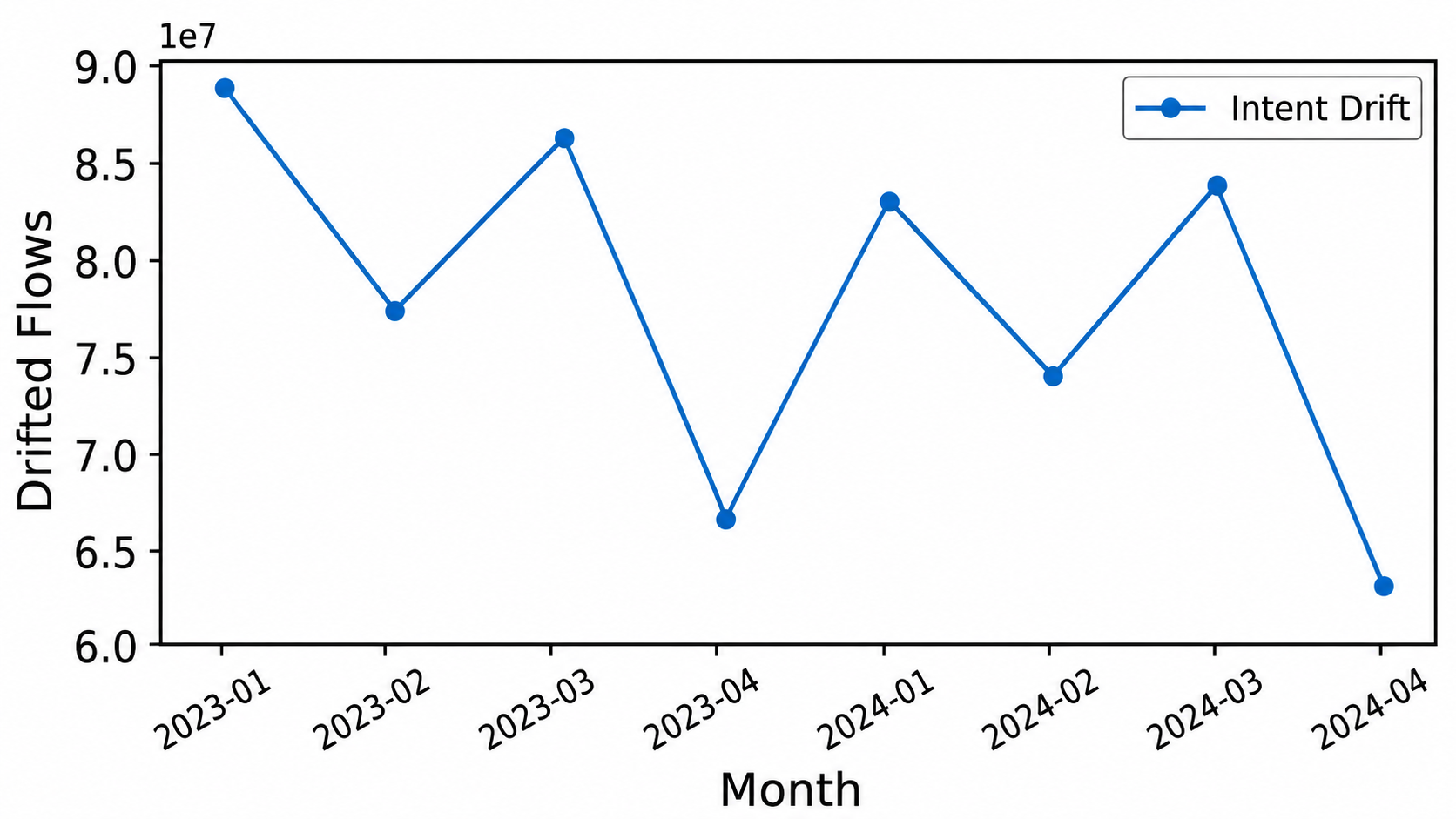}
    \caption{Monthly Intent Drift Across Intent Policies.}
    \label{fig:monthly_intent_drift}
    \vspace{-10pt}
\end{figure}

Figures~\ref{fig:monthly_policy_violations} and \ref{fig:monthly_intent_drift} establish the temporal stability of these metrics across matching operational windows in 2023 and 2024. While total volumes experience predictable seasonal fluctuations, peaking in January and March before declining in April, the structural policy hierarchy ($V_{\text{Strict}} > V_{\text{Balanced}} > V_{\text{Permissive}}$) remains immutable over time. The temporal variations in intent drift show  statistical independence from the enforced policy regimes, showcasing that underlying data-plane behavioral anomalies can propagate invariant to the security constraints defined by a high-level intents. 

%% file: 6DiscussionConclusion.tex
\begin{figure}[htbp]
    \centering
    \includegraphics[width=\linewidth]{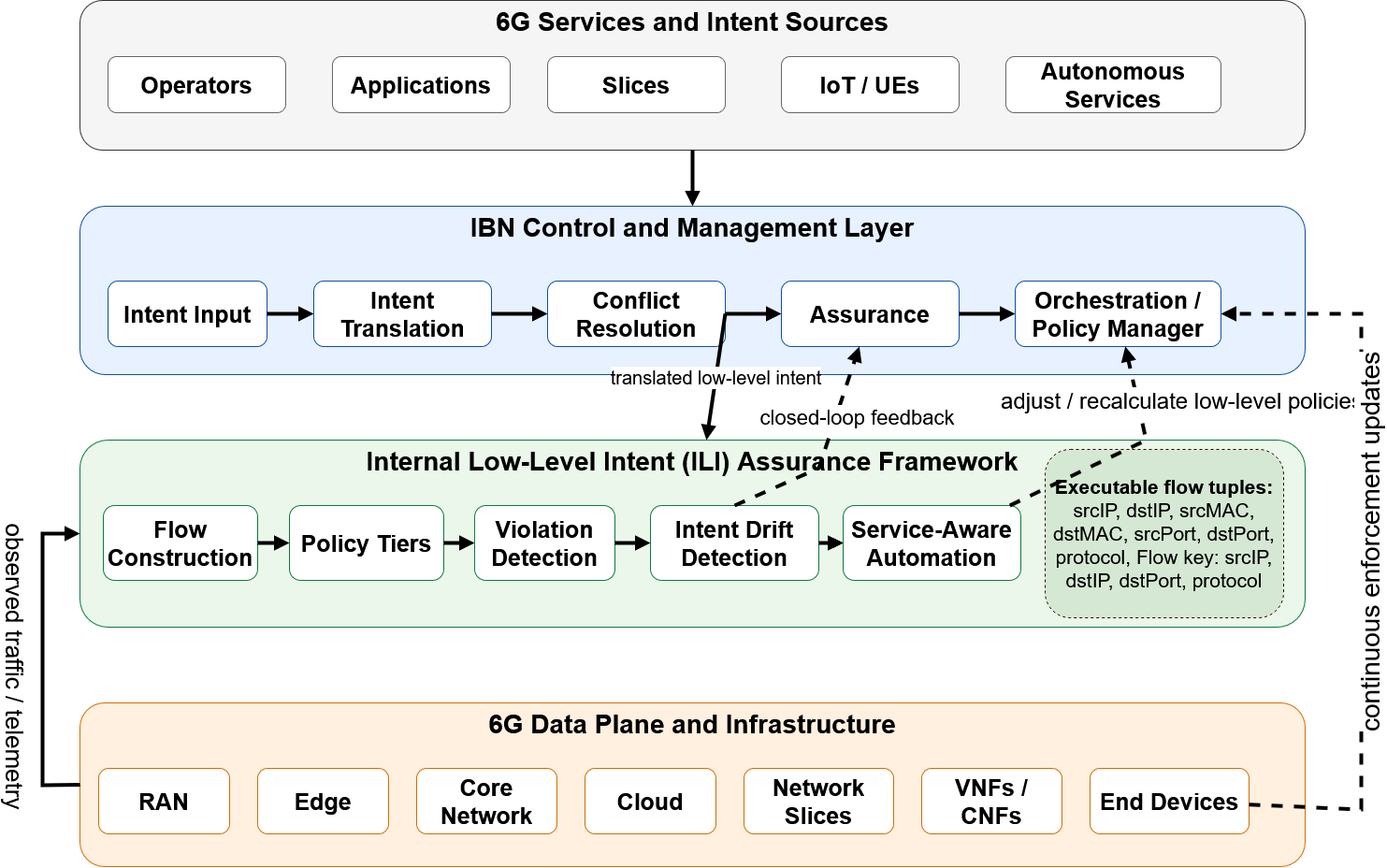}
    \caption{Closed-loop IBN orchestration architecture for autonomous 6G environments, detailing the top-down intent translation engine and bottom-up ILI telemetry verification.}
    \label{fig:6g_ibn_architecture}
    \vspace{-10pt}
\end{figure}
\section{6G Implications and Future Research Opportunities}
These empirical findings provide critical architectural insights for designing autonomous orchestration fabrics in production 6G environments. While traditional security mechanisms rely on bottom-up, reactive scripting to remediate anomalous traffic, demonstrating that structural variations in intent drift exhibit statistical independence from active policy regimes confirms that data-plane anomalies propagate invariant to high-level security constraints. Consequently, advanced Intent-Based Networking (IBN) frameworks must translate granular data-plane telemetry into top-down, semantic abstractions that mirror abstract behavioral expectations rather than static configuration states. Decoupled tracking of policy violations ($V_i$) and intent drift ($D$) via the proposed 7-tuple ILI interface equips closed-loop orchestrators with the telemetry primitives required to see past administrative masking. When anomalous flows are exposed beneath loose administrative rules, the orchestration engine leverages the Intent Behavioral Language (IBL) compiler to bypass configuration errors, dynamically recalculate multi-tenant policies, and push down updated configuration primitives to enforce immediate, behavior-defined operational boundaries across distributed 6G network slices. Moving forward, the conceptual and empirical contributions established in this study delineate three strategic, high-impact research opportunities for autonomous network governance and intent verification architectures.

This closed-loop orchestration cycle is architecturalized in Fig.~\ref{fig:6g_ibn_architecture}, which maps the top-down translation pipeline from high-level user intent down to line-rate data-plane execution elements. The framework demonstrates how the decoupled tracking of policy violations ($V_i$) and intent drift ($D$) via the Internal Low-Level Intent (ILI) interface explicitly feeds back into the orchestration engine, enabling automated, zero-trust policy synthesis and verification. By shifting the remediation loop from manual configurations to an autonomic feedback fabric, the system continuously ensures alignment between declarative IBL expectations and live physical network states.

\subsubsection{Dynamic Intent Re-Compilation \& Synthesis Engines} 
A primary challenge in closed-loop IBN is mitigating conflicting configurations during automated runtime remediation. Future work will investigate the continuous compilation mechanics required to translate decoupled ILI drift telemetry back into high-level declarative state changes. This requires designing mathematical synthesis models that dynamically update abstract security parameters without introducing semantic contradictions or violating structural invariants across 6G infrastructures~\cite{chowdhury2025framework}.

\subsubsection{Contextual Disambiguation} 
To scale the semantic fidelity of intent assurance in highly dynamic environments, the ILI interface must evolve beyond the static port classification sets evaluated in this study. Enriching ILI with multi-dimensional telemetry context, such as topological path profiles, destination network roles, protocol-specific state machine progression, and multi-temporal correlation patterns across concurrent communication streams, will allow the assurance fabric to accurately distinguish benign, seasonal operational drift from malicious, high-entropy behavioral anomalies.

\subsubsection{Line-Rate Ingestion and Distributed SmartNIC Execution} 
Transitioning from offline telemetry profiling to live, line-rate intent assurance demands extreme execution efficiency. Future architectures will focus on mapping the ILI verification pipeline directly onto programmable data planes, exploring distributed consensus and high-throughput algorithms to execute 7-tuple flow evaluations at terabit scale. Offloading this verification logic onto P4-enabled switches and SmartNICs will minimize end-to-end rule recalculation latencies and optimize hardware resource utilization during active zero-trust enforcement scenarios.
\section{Conclusion}
This paper presented a data-plane flow-level assurance framework for Intent-Based Networking (IBN) by formalizing raw network stream identifiers into standardized 7-tuple vectors, establishing an Internal Low-Level Intent (ILI) telemetry interface. Evaluated against a massive empirical dataset of 100.91 million live flow records, the proposed framework successfully exposed a fundamental validation gap within conventional verification methods. The data demonstrated that as administrative security postures expand from Strict to Permissive configurations, formal Policy Violations ($V$) systematically decrease, whereas the underlying data-plane Intent Drift ($D$) remains invariant across all operational setups. This serves to highlight that conventional, violation-centric assurance frameworks suffer from administrative masking, remaining blind to high-entropy behavioral shifts. By tracking violations and behavioral drift, this framework resolves the compliance blind spot, enabling closed-loop IBN orchestrators with the data-driven primitives to bypass administrative masking and automatically enforce more precise, behavior-defined operational boundaries for IBN in 6G infrastructures.

 \section*{Acknowledgment}
The research was partially supported by NSF Award Number 2319793, CICI: TCR program, IRIS: Instrumentation for Research and Inter-institutional SOC.